\documentclass[aps,pra,showpacs,preprint,superscriptaddress]{revtex4-1}
\usepackage{geometry}                

\usepackage{color}      
\usepackage{graphicx}
\usepackage{amssymb}
\usepackage{epstopdf}
\begin{document}

\title{Young's double slit interference pattern from a twisted beam. }
\author{Olivier Emile}
\affiliation{LPL,  URU 435 Universit\'{e} de Rennes I, 35042 Rennes Cedex, France.}
\altaffiliation{Corresponding author}
\email{olivier.emile@univ-rennes1.fr}
\author{Janine Emile}
\affiliation{IPR, UMR CNRS 6251, Universit\'{e} de Rennes I, 35042 Rennes Cedex, France.}
\date{\today}



\begin{abstract}

A wide range of diffractive elements have been used to evaluate the topological charge of Laguerre-Gaussian beams. Here we show theoretically and experimentally that this charge can be simply and readily measured from the interference pattern in Young's double slit experiment. It can be evaluated from the twisting order of the interference. The results are confronted with previously published studies. The potentialities of the method are then compared with existing techniques.

\end{abstract}

\pacs{42.50.Tx, 42.87.Bg, 42.50.Xa, 42.25.Hz}

\maketitle


\section{Introduction}
Young's double slit is may be one of the most popular and fascinating experiment in physics. As R. Feynman said: "It is a phenomenon which is impossible to explain in any classical way and which has in it the heart of quantum mechanics"  \cite{Feynman}. Curiously, when a phase difference is added in between the two interfering paths, like in an Aharonov-Bohm type experiment \cite{Peskin}, the interference pattern is shifted. However, along a single slit, the wave encounters exactly the same phase shift and thus the interference fringes remain straight lines. On the other hand, a new category of waves, called "twisted waves" appeared in the 1990s, which phase distribution is not uniform in a plane perpendicular to the direction of propagation  \cite{Bazhenov,Allen}. Since its first observation in optics, it has then found many applications in various domains including microwaves \cite{Thide}, atom optics \cite{Andersen}, quantum cryptography \cite{Leach}, telecommunications \cite{Wang,Tamburini0}, astronomy \cite{Tamburini}, biophysics \cite{Grier}, acoustics \cite{Demore} and electron beams \cite{Verbeeck}. Besides, these waves are known to produce diffraction or interference patterns as well \cite{Harris,Leach2,Sztul,Hickmann,Mei,Gao3,Belmonte,Emile}. However, researchers in the field mainly use  the diffraction by a triangle aperture \cite{Anderson,Mourka,Hickmann2}, or transformations with cylindrical or tilted spherical lens \cite{Denisenko,Vaity}, to characterize the beam whereas the double slit experiment is hardly ever implemented  although it seems easy to settle up and handle. The aim of this letter is to seek an analytical expression of the interference pattern of a Laguerre-Gaussian (LG) beam in a double slit experiment, to compare it with experimental observations and to investigate practical applications in the characterization of LG beams.

\section{Theoretical considerations}

\begin{figure}[htbp]
\includegraphics[width=5.5in]{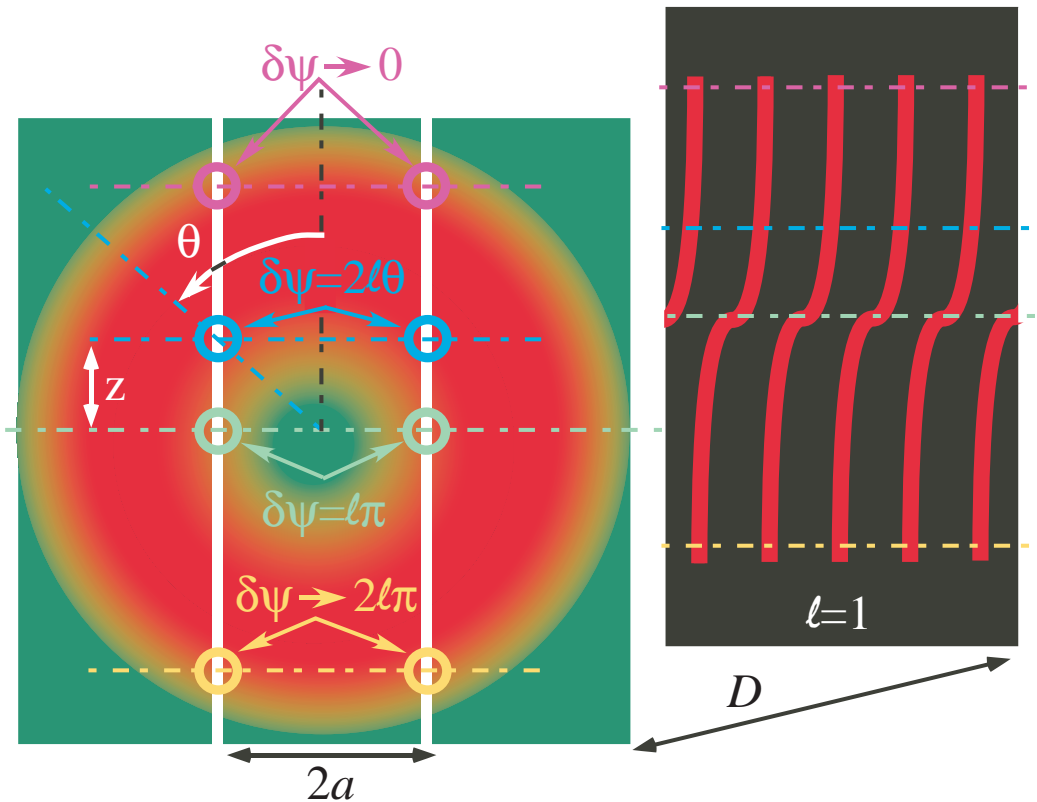}
\caption{\label{fig2} Schematic of the phase of the LG beam impinging on a double slit experiment. $2a$: distance between the slits, D distance between the slits and a screen. Two corresponding points at the same height have a phase difference equal to $\delta \psi = 2l\theta $. This leads to a twisted interference pattern.}
\end{figure}

From a theoretical point of view, the phase $\psi$ of the twisted beam, on a plane perpendicular to the direction of propagation, is not uniform as for usual plane waves. It varies from 0 to $2l\pi$ as one makes one complete turn around the direction of propagation \cite{Allen}. $l$ is called the Topological Charge (TC) of the beam. The phase writes $\psi(\theta) = l\theta$, where $\theta$ is the usual polar coordinate (see figure \ref{fig2}). $\theta$ is related to the coordinate $z$ along the vertical direction by the following relation: $\tan(\theta) = a/z$, where $2a$ is the distance between the slits. Then the phase difference $\delta \psi$ between the two paths of the double slit experiment, at a given height $z$, writes, $\delta \psi = \psi(\theta)-\psi(-\theta) = 2l\theta =2l\tan^{-1}(a/z)$. The intensity variation $I(x)$ due to the interference between the paths along the horizontal axis  can be written as \cite{Feynman}
\begin{equation}
I(x) = I_0\cos ^2 (2\pi x a / (\lambda D)+ \delta \psi) 
\label{EQ1}
\end{equation}
where $D$ is the distance between the slits and the screen used for observation. The interference pattern in the $x$ direction varies as 
\begin{equation}
2\pi x a / (\lambda D)+ 2l\tan^{-1}(a/z)
\label{EQ2}
\end{equation}

In particular, this means that as $z$ tends towards $+\infty$ (i.e. in the upper zone of the laser beam where $\theta=0$), the interference fringes should correspond to the usual pattern of a plane wave. The two interfering beams have the same phase on the double slit. Using exactly the same reasoning, as $z$ tends towards $-\infty$ (i.e. in the bottom zone of the laser beam where $\theta=\pi$) the interference fringes also correspond to the usual pattern of a plane wave. The phase difference equals $2l\pi$.  

Actually equations \ref{EQ1} and \ref{EQ2} offer a general description of the double slit interference with a LG beam. This theoretical result is in disagreement with what has been found previously \cite{Sztul}, where, from numerical calculations, a $\pi$ phase difference between the top and the bottom of the slits is predicted (see relation (4) and figure 2d \cite{Sztul}). For a correct description, one has to first take into account the phase difference $\delta \psi $ between the interfering paths at the top and at the bottom separately (see figure \ref{fig2}). Second, one has to compare the resulting $\delta \psi$ between the top and the bottom of the slits. It unambiguously equals $2\pi$. 

\section{Experiments and results}

\begin{figure}[htbp]
\includegraphics[width=5.5in]{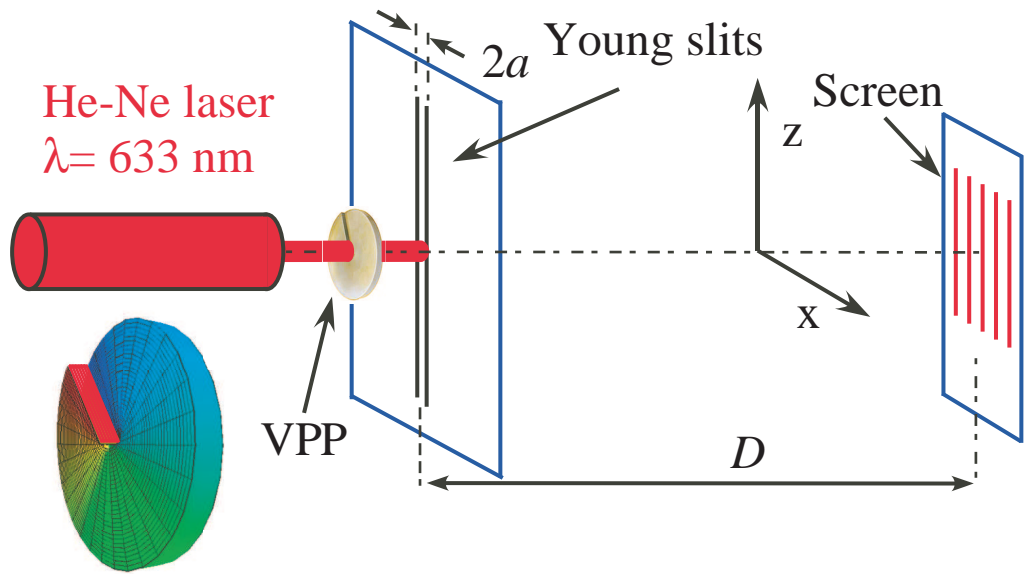}
\caption{\label{fig1} Experimental set up. 
VPP: Vortex Phase Plate. The zoom of the VPP shows a variable thickness that induces a phase variation in order to generate a LG beam.}
\end{figure}

From an experimental point of view, let us consider a typical double slit experiment like the one that can be found in textbooks \cite{Feynman}. We replace the usual light source by a twisted laser beam (see figure \ref{fig1}). The twisted beam is here a LG beam, but the experiment could be implemented for any twisted beam. It is generated from the fundamental beam of a red He-Ne laser ($\lambda=633$ nm, Melles Griot). The beam passes through a vortex phase plate \cite{Beijersbergen} (RC Photonics) with TC that can be chosen from $l = 0$ to $l = 3$ and a telescope (final beam waist 0.7 mm), before impinging on the double slit experiment. The slits are 3 cm long and $70~ \mu$m large. The distance between them is $2a = 300~ \mu$m. Pictures of the interference patterns are taken on a screen at a distance $D = 4$ m with a camera. 

\begin{figure}[htbp]
\includegraphics[width=5.5in]{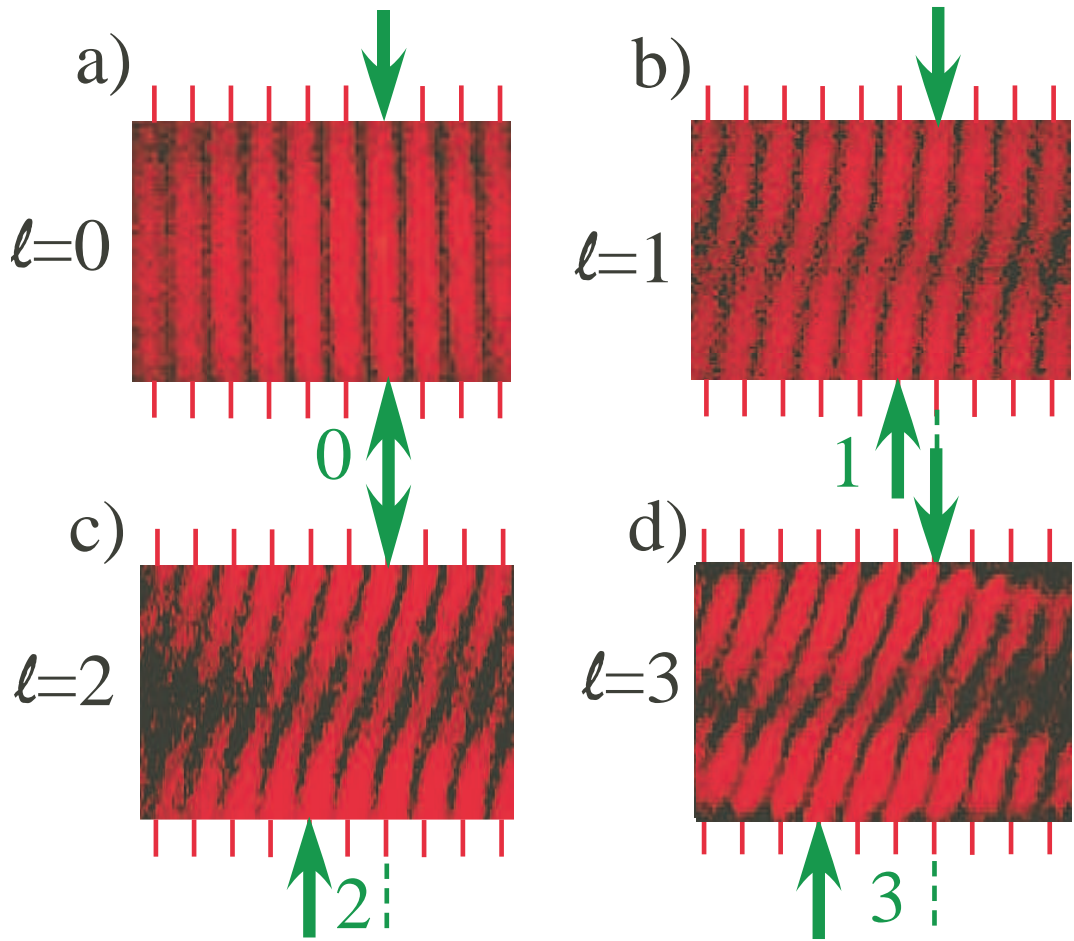}
\caption{\label{fig3} Twisted interference patterns for beams with a) $l = 0$; b) $ l = 1$; c) $ l = 2$; d) $ l = 3$. The twist corresponds to $l$ bright fringes of the interference pattern. The green arrows indicate the fringe order shift. }
\end{figure}

Figure \ref{fig3} presents several photographs of the interference pattern for various values of $l$. In figure  \ref{fig3}a ($l=0$), one recognizes the usual interference pattern of the typical double slit experiment using plane wave sources. In particular, the fringes are straight lines. The diffraction pattern has the same symmetry as the diffracting object. It has a cylindrical symmetry for a diffracting hole and has a Cartesian symmetry for diffracting slits \cite{Born}. However, for a twisted beam, for example for $l =1$ (see figure  \ref{fig3}b), the interference pattern is twisted. The fringes are not straight lines any more. They follow a  $2\tan^{-1}(a/z)$ variation, as expected. The interferences, at the bottom zone of the laser beam, have been shifted by exactly one interference order compared with the ones corresponding to the upper zone of the laser beam (see green arrows on the figure). This is a very unusual behaviour of the interference pattern in a double slit experiment. Moreover, as $l$ is further increased, the twist of the fringes becomes more and more important. For $l = 2$, the fringes at the bottom zone (see figure  \ref{fig3}c) have been shifted by two interference orders compared to the ones at the upper zone, whereas for $l = 3$ (see figure  \ref{fig3}d), they correspond to three interference orders. Analogously, for $l = n$, they should correspond to $n$ orders. 

 \section{Discussion}
 
Actually this experiment is a quick and easy to handle way to determine the TC of a twisted beam. The twist of the fringes can be readily seen on a screen with the naked eye. One has only to count the number of twists from the top of the interference pattern to the bottom, following the whole interference pattern. However, one cannot determine the TC of the beam when considering the top and the bottom of the fringes only, since the bottom fringes are shifted exactly by an integer number order of fringes. Besides, in order to obtain such patterns, there must be some light impinging on the various zones of the slits. In particular, since the center of the laser beam is a vortex, one has to adapt the distance $2a$ between the slits to obtain a good and usable pattern. 

Let us be more quantitative and define a minimum and a maximum distance $2a$ between the slits to determine the TC of the beam. There are two criteria concerning: (i) the visibility of the fringes, (ii) the zone of the beam probed. For (i), from the naked eye, the intensity of a fringe should not vary more than a factor of 5  to be clearly detected. If one considers a $l=1$ LG beam with a waist $w$, the beam intensity $I(r)$ ($r$ being the polar coordinate) is proportional to

\begin{equation}
I(r) \sim \frac{r^2}{w^2} \exp{\frac{-2r^2}{w^2} } 
\label{E2}
\end{equation}

The  beam intensity is maximum for $r=w\sqrt{l}/\sqrt{2}$. The intensity of a single fringe $F(z)$ follows the same variation as $I(r)$, taking $r=\sqrt{a^2+z^2}$. The minimum intensity of the fringe is for $z=0$ ($\theta=\pi/2$). The minimum distance $2a$ between the slits, corresponding to $F_0/5$  is $2w/5$. The maximum intensity $F_0$ corresponds to $z=0.95w/\sqrt{2}$. As $z$ increases further, the intensity of the fringe decreases and the intensity $F_0/5$ corresponds to $z=2w/\sqrt{2}$ ($\theta=0.14$ rad). The phase variation of the beam is thus probed in the region $0.14 < \theta < \pi -0.14$. Let us move to the (ii) criterium. We assume that the TC of the beam could be easily identified if we probe more than $80\%$ of the beam, i.e. $0.35<\theta<\pi-0.35$. Then criterium (i), applied to $\theta=0.35$ rad, leads to $a < w/2$. Thus, practically, the distance $2a$ between the slits should be such as $w/5 < a < w/2$ in order the TC of the LG beam to be easily determined. In our case, $w=700~ \mu$m and $a = 150~ \mu$m, fulfill the two criteria. Besides, for $a=w/5$, the height of the slits should be at least more than $4w/\sqrt{2}\simeq3w$ ($\theta=0.14$ rad) not to truncate the beam. This last argument about the height of the slits may be the reason why previous works incorrectly evidence a smaller twist of the fringes \cite{Sztul}. Finally, the beam must be centered on the middle of the slits. 

\begin{figure}[htbp]
\includegraphics[width=5.5in]{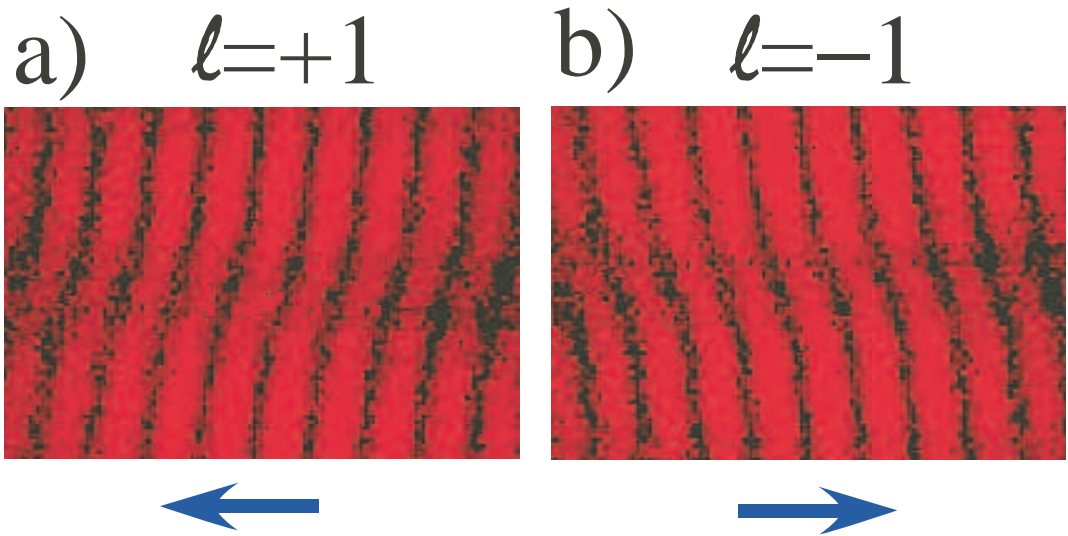}
\caption{\label{fig4} Twisted interference patterns for two opposite signs of $l$. The interferences are shifted in opposite directions (see blue arrows). }
\end{figure}

So far, we have shown that the interference pattern of a double slit experiment allows to precisely determine the TC of a twisted beam. One may thus wonder whether the sign of this charge could be also fixed. Actually, following equation \ref{EQ2}, the sign of the twist of the pattern should be reversed when changing the sign of the TC. Experimentally, let us reverse the orientation of the vortex phase plate (see figure \ref{fig1}) so as to reverse the sign of $l$. Figure~\ref{fig4} shows the interference pattern for $l = 1$ and $l = -1$. The sign of the twist of the fringes is reversed, as expected. It can be determined unambiguously. Could this technique be used to precisely determine non integer TC \cite{Basisty,Lee,Tamburini2}? It seems difficult since the light intensity distribution is non symmetrical. This would scramble the interference pattern. Nevertheless, when the slits are aligned with the light discontinuity, the symmetry is restored and the non integer TC could be estimated. 

Several interferential techniques have been proposed to determine the TC of the beam. The most popular one is perhaps the diffraction by a triangular aperture \cite{Hickmann,Anderson,Mourka,Hickmann2} that needs to count the number of diffracted spots. However, for $l>7$ it is sometimes hard to determine the number of spots. There are similar techniques using more complicated apertures such as hexagonal aperture \cite{Liu}, annular aperture  \cite{Guo2,Tao}, or multi points interferometer \cite{Guo}. One can mention the interference of the LG beam with itself that needs a biprism and a lens \cite{Harris,Belmonte}, the diffraction by an edge \cite{Cui} and by a single slit \cite{Ferreira}, or even with a double angular slit \cite{Liu2}, where the relationship between the pattern and $l$ is not straightforward. The image in the focal plane of a cylindrical lens \cite{Denisenko} or equivalently with a tilted spherical lens  \cite{Vaity} has also been performed. Nevertheless, Young's double slit experiment is indeed easy to settle and use. The measurements could be performed with the naked eye. Even with high TC, or with partially coherent beams \cite{Zhao} or with tiny spots \cite{Emile} the experiment would be of practical use. The calculations are elementary and the interpretation of the experimental results is straightforward. Besides, the sign of the TC could be easily determined.

\section{Conclusion}
 This easy-to-settle experiment enables to precisely measure, with the naked eye, the value and the sign of the topological charge of a twisted beam. These results are in very good agreement with a simple model, leading to analytical expressions. This model is indeed a comprehensive description of Young's double slit experiment with twisted beams. Since this experiment could be performed for any kind of waves in each specific domain (optics, radio electromagnetic waves, acoustics, particle beams such as electron beams, ...), this  procedure can be easily implemented to determine the characteristics of the beam, when dealing with twisted beams. This could even be performed  with X-ray with newly generated vortices \cite{Kohmura}. This could also find applications in the growing field of light communication in the sorting of the multiplexed twisted beam \cite{Wang,Gatto} or in encoding data for entangle purposes \cite{Pan}.
\section*{Acknowledgements}
We acknowledge support from the Universit\'{e} de Rennes 1 via a d\'{e}fi \'{e}mergent action, technical support from J.R. Th\'{e}bault, and valuable help from A. Voisin. 	












\begin{thebibliography}{39}

\bibitem{Feynman}
R.P.~Feynman, R.~Leighton, and M.~Sands,
\textit{The Feynman lecture on physics.} (Addison-Wesley, Massachusetts, 1965).

\bibitem{Peskin}
M.~Peskin and A.Tonomura,
\textit{The Aharonov-Bohm effect.} (Springer-Verlag, Berlin, 1989).

\bibitem{Bazhenov}
V.Y.~Bazhenov, M.V.~Vasnetsov, and M.S.~Soskin,
JETP Lett. \textbf{52} (1990) 429.

\bibitem{Allen}
L.~Allen, M.W.~Beijersbergen, R.J.C.~Spreeuw, and J.P.~Woerdman,
Phys. Rev. A \textbf{45} (1992) 8185 .

\bibitem{Thide}
B.~Thid\'{e}, H.~Then, J.~Sj\"{o}holm, K.~Palmer, J.~Bergman, T.D.~Carrozzi, Y.N.~Istomin, N.H.~Ibragimov, and R.~Khanitova,
Phys. Rev. Lett. \textbf{99} (2007) 087701.

\bibitem{Andersen}
M.F.~Andersen, C.~Ryu, P.~Clad\'{e}, V.~Natarajan, A.~Vaziri, K.~Helmerson, and W.D.~Philipps,
Phys. Rev. Lett. \textbf{97} (2006) 170406.

\bibitem{Leach}
J.~Leach, B.~Jack, J.~Romero, A.K.~Jha, A.M.~Yao, S.~Franke-Arnold, D.G.~Ireland, R.W.~Boyd, S.M.~Barnett, and M.J.~Padgett,
Science \textbf{329} (2010) 662.

\bibitem{Wang}
J.~Wang, J.Y.~Yang, I.M.~Fazal, N.~Ahmed, Y.~Yan, H.~Huang, Y.~Ren, Y.~Yue, S.~Dolinar, M.~Tur, and A.E.~Willner,
Nat. Photonics \textbf{6} (2012) 488.

\bibitem{Tamburini0}
F.~Tamburini, E.~Mari, A.~Sponselli, B.~Thid\'{e}, A.~Bianchini, and F.~Romato,
New J. Phys. \textbf{14} (2012) 033001.

\bibitem{Tamburini}
F.~Tamburini, B.~Thid\'{e}, G.~Molina-Terriza, and G.~Anzolin,
Nature Phys. \textbf{7} (2011) 195.

\bibitem{Grier}
D.G.~Grier,
Nature \textbf{424} (2003) 810.

\bibitem{Demore}
C.E.M.~Demore, Z.~Yang, A.~Volovick, S.~Cochran, M.P.~MacDonald, and G.C.~Spalding,
Phys. Rev. Lett. \textbf{108} (2012) 194301.

\bibitem{Verbeeck}
J.~Verbeeck, H.~Tian, and P.~Schattschneider,
Nature \textbf{467} (2010) 301.

\bibitem{Harris}
M.~Harris, C.A.~Hill, and J.M.~Vaughan,
Opt. Commun. \textbf{106} (1994) 161.

\bibitem{Leach2}
J.~Leach, J.~Courtial, K.~Skeldon, S.M.~Barnett, S.~Franke-Arnold, and M.J.~Padgett,
Phys. Rev. Lett. \textbf{92} (2004) 013601.

\bibitem{Sztul}
H.I.~Sztul and R.R.~Alfano,
Opt. Lett. \textbf{31} (2006) 999.

\bibitem{Hickmann}
J.M.~Hickmann, E.J.S.~Fonseca, W.C.~Soares, and S.~Chavez-Cerda,
Phys. Rev. Lett. \textbf{105} (2010) 053904.

\bibitem{Mei}
Z.~Mei and J.~Gu,
Appl. Phys. B \textbf{99} (2010) 571.

\bibitem{Gao3}
C. Gao, X. Qi, Y. Liu, J. Xin, and L. Wang, 
Opt. Commun. \textbf{284} (2011) 48.

\bibitem{Belmonte}
A.~Belmonte and J.P.~Torres,
Opt. Lett \textbf{37} (2012) 2940.

\bibitem{Emile}
O.~Emile, A.~Voisin, R.~Niemiec, B.~Viaris de Lesegno, L.~Pruvost, G.~Ropars, J.~Emile, and C.~Brousseau,
EPL \textbf{101} (2013) 54005. 

\bibitem{Anderson}
M.E.~Anderson, H.~Bigman, L.E.~de Araujo, and J.L.~Chaloupka, 
J. Opt. Soc. Am. B \textbf{29} (2012) 1968.

\bibitem{Mourka}
A.~Mourka, J.~Baumgarti, C.~Shanor, K.~ Dholakia,  and E.M.~Wright,
Opt. Express \textbf{19} (2011) 5760.

\bibitem{Hickmann2}
J.M.~Hickmann, E.J.S.~Fonseca, and A.J.~Jesus-Silva,
EPL \textbf{96} (2011) 64006.   

\bibitem{Denisenko}
V.~Denisenko, V.~Shvedov, A.S.~Desyatnikov, D.N.~Neshev, W.~Krolikowski, A.~Volyar, M.~Soskin, and Y.S.~Kivshar,
Opt. Express \textbf{17} (2009) 23374. 

\bibitem{Vaity}
P.~Vaity, J.~Banerji, and R.P.~Singh,
Phys. Lett. A \textbf{377} (2013) 1154.

\bibitem{Beijersbergen}
M.W.~Beijersbergen, R.P.C.~Coerwinkel, M.~Kristensen, and J.P.~Woerdman 
Opt. Commun. \textbf{112} (1994) 321.

\bibitem{Born}
M.~Born and E.~Wolf,
\textit{Principle of optics. 7th Ed.} (Cambridge University Press, Cambridge, 1999).

\bibitem{Basisty}
L.V.~Basistiy, V.A.~Pas'ko, V.V.~Slyusar, M.V.~Soskin, and M.V.~Vasnetsov,
J. Opt. A: Pure Appl. Opt. \textbf{6} (2004) S166.

\bibitem{Lee}
W.M.~Lee, X.-C.~Yani, and K.~Dholakia,
Opt. Commun.  \textbf{239} (2004) 129.

\bibitem{Tamburini2}
F.~Tamburini, E.~Mari, B.~Thid\'{e}, C.~Barbieri, and F.~Romato,
Appl. Phys. Lett. \textbf{99} (2011) 204102.

\bibitem{Liu}
Y.~Liu and J.~Pu,
Opt. Commun. \textbf{284} (2011) 2424.

\bibitem{Guo2}
C.S.~Guo, L.L.~Lu, and H.T.~Wang,
Opt. Lett. \textbf{34} (2009) 3686.

\bibitem{Tao}
H.~Tao, Y.~Liu, Z.~Chen, and J.~Pu,
Appl. Phys. B \textbf{106} (2012) 927.

\bibitem{Guo}
C.S.~Guo, S.J.~Yue, and G.S.~Wei,
Appl. Phys. Lett. \textbf{94} (2009) 231104.

\bibitem{Cui}
H.X.~Cui, X.L.~Wang, B.~Gu, Y.N.~Li, J.~Chen, and H.~Wang,
J. Opt. \textbf{14} (2012) 055707.

\bibitem{Ferreira}
Q.S.~Ferreira, A.J.~Jesus-Silva, E.J.S.~Fonseca, and J.M.~Hickmann,
Opt. Lett. \textbf{36} (2011) 3106.

\bibitem{Liu2}
R.~Liu, J.~Long, F.~Wang, Y.~Wang, P.~Zhang, H.~Gao, and F.~Li,
J. Opt. \textbf{15} (2013) 125712.

\bibitem{Zhao}
C.~Zhao, Y.~Dong, Y.~Wang, F.~Wang, Y.~Zhang, and Y.~Cai,
Appl. Phys. B \textbf{109} (2010) 345.

\bibitem{Kohmura}
Y.~Kohmura, K.~Sawada, M.~Taguchi, T.~Ishikawa, T.~Ohigashi, and Y.~Suzuki,
Appl. Phys. Lett. \textbf{94} (2009) 101112.

\bibitem{Gatto}
A.~Gatto, M.~Tacca, P.~Martelli, P.~Boffi, and M.~Martinelli,
J. Opt. \textbf{13} (2011) 064018.

\bibitem{Pan}
J.W.~Pan, Z.B.~Chen, C.-Y.~Lu, H.~Weinfurter, A.~Zeilinger, and M.~Zukowski,
Rev. Mod. Phys. \textbf{84} (2012) 777. 777.

 
\end{thebibliography}
\end{document}